\documentclass[letterpaper]{article} 
\usepackage{aaai2027}  
\usepackage[hyphens]{url}  
\usepackage{amsmath}
\usepackage{graphicx} 
\usepackage{natbib}  
\usepackage{caption} 
\usepackage{algorithm}
\usepackage{algorithmic}
\usepackage{booktabs,multirow,makecell,graphicx}
\usepackage{multirow}
\usepackage{newfloat}
\usepackage{listings}
\DeclareCaptionStyle{ruled}{labelfont=normalfont,labelsep=colon,strut=off} 
\floatstyle{ruled}
\newfloat{listing}{tb}{lst}{}
\floatname{listing}{Listing}

\usepackage{booktabs}
\usepackage{caption}

\nocopyright 

\title{Omni-LiveAvatar: Minute-Level Real-Time Streaming Joint Audio-Video Avatar Generation}
\author{
    Lunjie Zhu\textsuperscript{\rm 1,\rm 2},
    Xingtong Ge\textsuperscript{\rm 1,\rm 2},
    Fangyu Lin\textsuperscript{\rm 1,\rm 2},
    Yi Zhang\textsuperscript{\rm 2},
    Zhening Liu\textsuperscript{\rm 1},
    Mengfei Li\textsuperscript{\rm 1,\rm 2},\\
    Yumeng Zhang\textsuperscript{\rm 1},
    Guanglu Song\textsuperscript{\rm 2},
    Yu Liu\textsuperscript{\rm 2},
    Jun Zhang\textsuperscript{\rm 1}\corresponding
}

\affiliations{
    \textsuperscript{\rm 1}iComAI Lab, Hong Kong University of Science and Technology \\
    \textsuperscript{\rm 2}Vivix Group Limited\\
    lzhubb@connect.ust.hk,
    eejzhang@ust.hk
}

\begin{document}


\maketitle

\begin{figure*}[t]
    \centering
    \includegraphics[width=\textwidth]{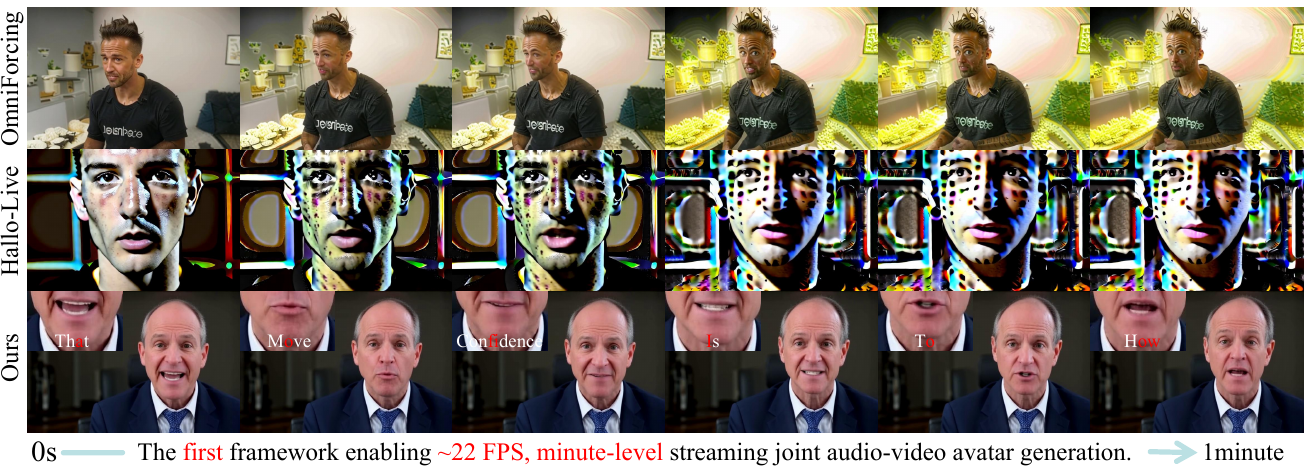}
    \caption{Qualitative comparison of minute-level generation.
    Compared with state-of-the-art baselines, Omni-LiveAvatar preserves
    consistent avatar and background appearance, and maintains tight audio-video
    alignment over a minute-long rollout, whereas OmniForcing and Hallo-Live
    exhibit severe appearance drift. In the bottom row, the syllable uttered
    at each frame is highlighted in red, and the insets show the corresponding
    mouth shapes for lip-audio alignment.}
    \label{fig:teaser}
    \vspace{-10pt}
\end{figure*}


\begin{abstract}
Joint audio-video generative models serve as foundation for immersive and interactive digital-human generation. Nevertheless, most existing models rely on bidirectional attention and multi-step
denoising and can generate only short clips, making them unsuitable for real-time interaction over extended durations. We present \textbf{Omni-LiveAvatar}, the first framework for minute-level, real-time streaming joint audio-video avatar generation. Specifically, we propose (1) a \textit{progressive autoregressive distillation pipeline} that transfers a large bidirectional joint audio-video diffusion model into a few-step autoregressive generator without auxiliary stabilization mechanisms; (2) a \textit{synchronized audio-video long-short-term memory} that preserves global consistency under a bounded memory budget; and (3) a \textit{hierarchical rolling prompt planning} strategy that enables coherent semantic evolution and seamless prompt transitions. Extensive experiments show that Omni-LiveAvatar generates high-quality, synchronized minute-level avatars in real time. In terms of speed, it achieves a \textbf{33$\times$} generation speedup over its teacher, LTX-2, on a single NVIDIA H200 GPU; in terms of generation quality, it outperforms accelerated baselines across visual quality, audio quality, cross-modal synchronization, and human fidelity. Our code is available at \url{https://github.com/Aoko955/Omni-LiveAvatar}
\end{abstract}

\section{Introduction}

Recently, remarkable progress has been made in joint audio-video generative models~\citep{LTX-2,ovi} for synthesizing high-fidelity, synchronized multimodal content. These advances lay a strong foundation for interactive digital-human applications
that demand realistic appearance, coherent motion, and accurate lip-synced speech~\citep{PrismMirror}.  However, most existing models rely on bidirectional attention and multi-step
denoising, which incurs high inference latency and confines generation to short clips. This makes them unsuitable for latency-sensitive applications that require continuous interaction over extended durations.

To enable real-time streaming generation, recent works~\cite{OmniForcing,hallo-live} adapt autoregressive distillation techniques~\citep{senseflow,salt} originally developed for video generation to joint audio-video models. However, these methods largely rely on modality-specific remedies, such as audio sink tokens or additional future audio context, and external components such as reward
models. While effective, they do not examine whether the classic distillation pipeline remains suitable for joint generation given the increased model scale and the inherent complexity of multimodal modeling. Moreover, existing joint audio-video autoregressive distillation methods are primarily restricted to short-clip generation, leaving stable minute-level streaming avatar generation underexplored.

To address these challenges, we present \textbf{Omni-LiveAvatar}, the first framework for minute-level, real-time streaming joint audio-video avatar generation. Rather than patching an unstable training pipeline with modality-specific remedies, we revisit the classic distillation pipeline at a more fundamental level and develop a \emph{progressive autoregressive distillation} pipeline that converts a bidirectional teacher into a real-time streaming generator without auxiliary designs. 
For minute-level generation, we introduce a \emph{synchronized audio-video long-short-term memory} mechanism that combines periodically re-anchored long-term memory with a rolling key-value (KV) cache, preserving global consistency and recent context under a bounded memory budget. To schedule long prompts under the joint audio-video rolling-forcing~\citep{rollingforcing} setting without semantic conflicts across prompt transitions, we propose a \emph{hierarchical rolling prompt planning} strategy that decomposes the full prompt into a fixed global prompt, which specifies persistent appearance, background, and overall motion, as well as block-level local prompts that advance together with the rolling window, thereby avoiding the overlapping or blurred speech caused by abrupt prompt switching.

In summary, our work focuses on real-time, long-duration joint audio-video avatar synthesis and makes the following contributions:

\begin{itemize}
    \item We propose a progressive autoregressive distillation pipeline that transfers a large bidirectional audio-video diffusion model into a few-step causal generator without auxiliary stabilization mechanisms.

    \item We design a synchronized audio-video long-short-term memory mechanism that sustains global consistency and cross-modal alignment under a bounded memory budget.

    \item We introduce a hierarchical rolling prompt planning strategy that enables coherent semantic transitions under the joint rolling-forcing setting.

    \item Extensive experiments show that Omni-LiveAvatar generates high-quality, synchronized minute-level audio-video avatars in real time, achieving a \textbf{33$\times$} generation speedup over its teacher LTX-2 while outperforming accelerated baselines across all evaluated dimensions.
\end{itemize}

\section{Related Work}

\paragraph{Audio-Driven Avatar Generation.}
Building on powerful diffusion models~\citep{wan,hunyuanvideo}, audio-driven avatar generation has witnessed remarkable advances in visual fidelity~\citep{emo,hallo,echomimic,hallo3,hunyuanvideo-avatar}, long-duration consistency~\citep{loopy,hallo2,infinitetalk,stableavatr}, and real-time streaming capability~\citep{vasa1,teller,liveavatar,soulxflashtalk}. However, these audio-driven methods rely on pre-generated audios and lack mutual adaptation between the audio and video streams, complicating the overall generation pipeline and limiting flexible audio-video coordination.


\paragraph{Autoregressive Joint Audio-Video Generation.}
Joint audio-video generation models produce both audio and video within a unified framework and generally follow three paradigms: cascaded generation~\citep{mmdit}, dual-stream modeling with cross-modal interaction~\citep{ovi,LTX-2}, and unified single-stream modeling~\citep{davinci-magihuman-2026}. However, they all rely on bidirectional attention and multi-step denoising, resulting in slow inference and high computational costs~\citep{ovi,LTX-2,remedygs,linvideo,qvgen}.

To enable real-time interactive applications, recent works adapt autoregressive distillation pipelines and acceleration techniques~\citep{causvid,selfforcing,flash-vaed} developed for video models to joint audio-video models. OmniForcing~\citep{OmniForcing} introduces audio sink tokens to stabilize joint audio-video causalization, while Hallo-Live~\citep{hallo-live} employs future-expanding audio attention and preference-guided distribution matching distillation (DMD) with external reward models. However, these approaches address the challenges of joint audio-video autoregressive distillation mainly through auxiliary stabilization mechanisms or external supervision, rather than fundamentally revisiting the distillation framework. Moreover, existing methods are designed and evaluated primarily for short-clip generation, leaving stable minute-level generation underexplored.

\section{Method}

\begin{figure*}[!t]
    \centering
    \includegraphics[width=\linewidth]{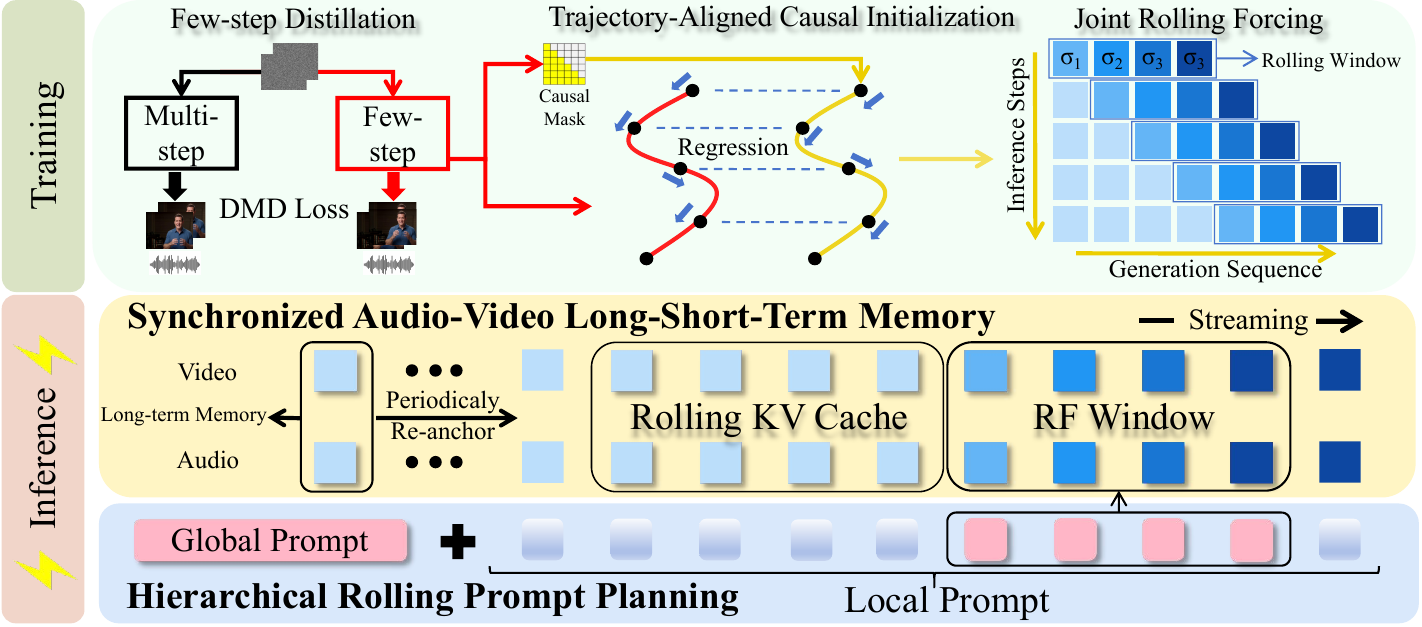}
    \caption{Overview of Omni-LiveAvatar.
The proposed progressive autoregressive distillation pipeline converts a large bidirectional audio-video diffusion model into a few-step causal generator without any auxiliary stabilization mechanism (top).
At inference, the synchronized audio-video long-short-term memory preserves global consistency while retaining recent context within a bounded memory budget for minute-level streaming inference (middle), and the hierarchical rolling prompt planning organizes global and local prompts for smooth long-form semantic evolution (bottom).}
    \label{fig:overview}
\end{figure*}

\subsection{Problem Formulation and Method Overview}
Given a text prompt $\mathcal{C}$ describing an avatar, scene, and audio content, our goal is to generate a minute-level avatar video $\mathbf{V}$ together with synchronized audio $\mathbf{A}$ in real time. While a natural direction is to extend autoregressive distillation methods from video generation to joint audio-video generation, this extension faces three challenges: (i) scaling the distillation to larger multi-modal models, where audio modeling and cross-modal coupling increase distillation difficulty; (ii) controlling cross-modal drift, where visual and audio drifts accumulate over time and reinforce each other, degrading identity consistency, speech quality, and lip synchronization; and (iii) scheduling heterogeneous semantics, to coordinate the slowly varying visual context with rapidly changing speech content over minute-long generation. Omni-LiveAvatar addresses these challenges with progressive autoregressive distillation, synchronized audio-video long-short-term memory, and hierarchical rolling prompt planning, whose details are presented in the following subsections.

\subsection{Progressive Autoregressive Distillation for Joint Audio-Visual Generation}

\begin{figure}[t]
    \centering
    \includegraphics[width=\linewidth]{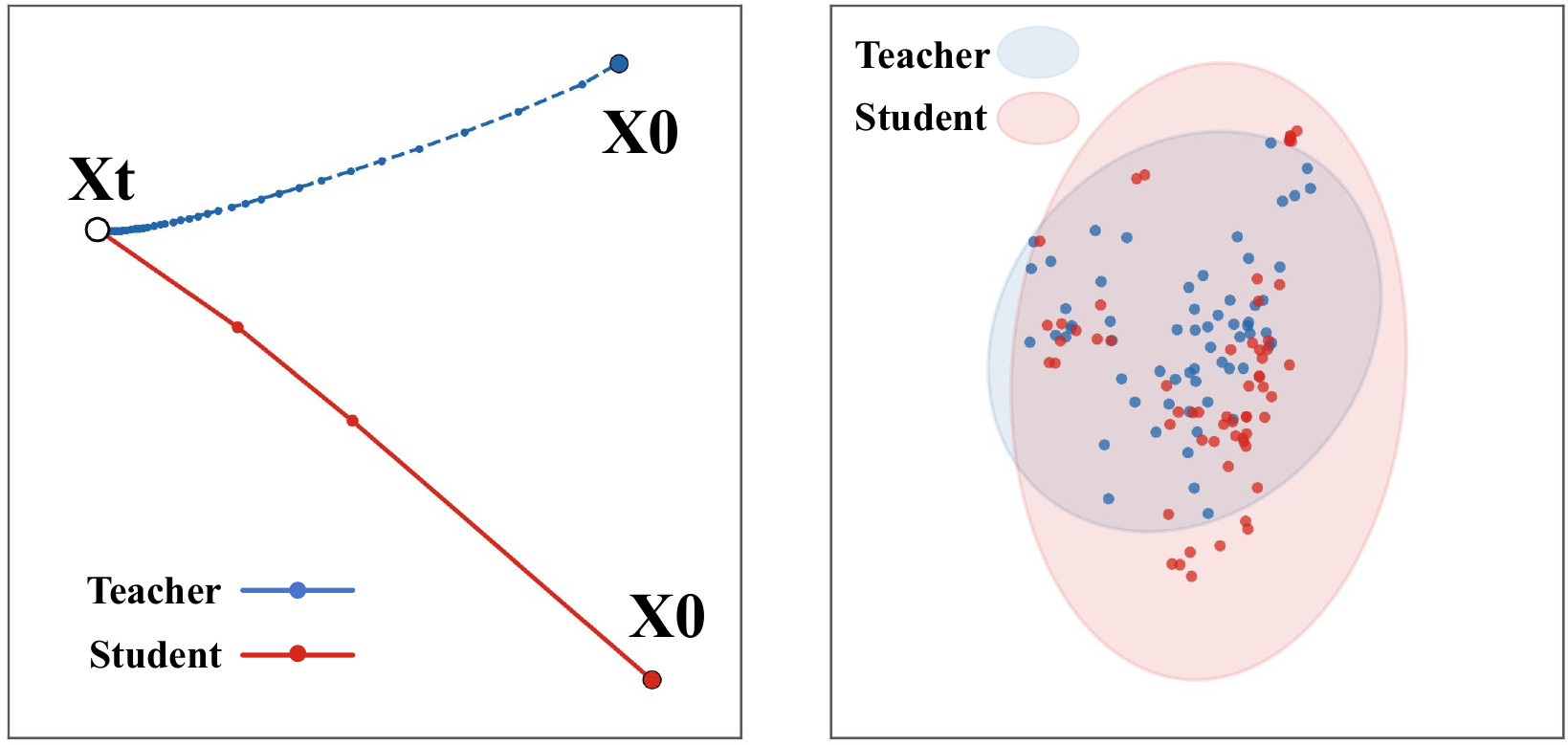}
    \caption{Visualization of the denoising trajectories~(left) and output distributions~(right) of the multi-step teacher and few-step student. Projected via principal component analysis, the visualization shows that DMD aligns the output distribution but disrupts the pointwise denoising trajectory.}
    \label{fig:dmd_trajectory}
\end{figure}

Existing autoregressive distillation methods for video generation~\citep{rollingforcing,selfforcing} typically initialize a few-step causal student by regressing the ordinary differential equation (ODE) trajectories of a pre-trianed bidirectional teacher. However, such initialization does not readily extend to joint audio-video generation, where the larger model scale, complex cross-modal coupling, and modality asymmetry make bidirectional-to-causal adaptation substantially more difficult. To obtain a stronger few-step causal initialization, OmniForcing~\citep{OmniForcing} decouples few-step distillation from causal adaptation by first applying distribution matching distillation (DMD) to obtain a lightweight few-step generator and then performing causal ODE regression. However, its ODE pairs are generated by the original multi-step teacher\protect\footnotemark[1]. As illustrated in Figure~\ref{fig:dmd_trajectory}, DMD matches the output distribution but does not preserve the pointwise denoising trajectory. Therefore, the causal student is forced to regress toward ODE targets from a different trajectory, which leads to unstable causal training and explains why OmniForcing requires additional stabilization mechanisms.

To address these issues, we develop a progressive autoregressive distillation pipeline comprising few-step distillation, trajectory-aligned causal initialization, and joint rolling forcing, enabling stable causal adaptation and efficient few-step generation without relying on auxiliary designs. Specifically, in Stage I, we first distill the pretrained bidirectional audio-video diffusion model \(G_{\mathrm{base}}^{\mathrm{bi}}\) into a bidirectional few-step generator \(G_{\mathrm{few}}^{\mathrm{bi}}\), optimizing the student with a joint audio-video DMD loss~\citep{dmd}:
\begin{equation}
\mathcal{L}_{\mathrm{Bi\text{-}DMD}}
=
\lambda_v\mathcal{L}_{\mathrm{DMD}}^v
+
\lambda_a\mathcal{L}_{\mathrm{DMD}}^a,
\end{equation}
where \(\lambda_v\) and \(\lambda_a\) balance the video and audio losses.

\footnotetext[1]{https://github.com/OmniForcing/OmniForcing/issues/8}

In Stage II, we convert \(G_{\mathrm{few}}^{\mathrm{bi}}\) into a causal few-step generator \(G_{\mathrm{few}}^{\mathrm{causal}}\) by applying a block-causal attention mask over audio-video macro-blocks, where \(B_k=(V_k,A_k)\) denotes the \(k\)-th macro-block containing temporally aligned video and audio latents over a one-second interval~\citep{OmniForcing}. To avoid trajectory mismatch and stabilize causal initialization, we generate ODE pairs with the frozen \(G_{\mathrm{few}}^{\mathrm{bi}}\) and train \(G_{\mathrm{few}}^{\mathrm{causal}}\) via ODE regression:
\begin{equation}
\begin{aligned}
\mathcal{L}_{\mathrm{C\text{-}ODE}}
=
&\lambda_v
\big\|
v_\theta^v(\mathbf{x}_t)
-
v_\phi^v(\mathbf{x}_t)
\big\|_2^2
\\
&+
\lambda_a
\big\|
v_\theta^a(\mathbf{x}_t)
-
v_\phi^a(\mathbf{x}_t)
\big\|_2^2 ,
\end{aligned}
\end{equation}
where \(\mathbf{x}_t\) is the joint noisy latent at denoising timestep \(t\), \(v_\phi\) and \(v_\theta\) denote the velocity predictions of the bidirectional few-step teacher \(G_{\mathrm{few}}^{\mathrm{bi}}\) and the causal student \(G_{\mathrm{few}}^{\mathrm{causal}}\) at \(\mathbf{x}_t\), respectively. \(\lambda_v\) and \(\lambda_a\) balance the video and audio regression losses.

However, strict block-causal rollout causes errors to propagate and accumulate unidirectionally, leading to severe drift during long-horizon inference. To address this, in Stage III, we introduce joint rolling forcing. Specifically, we construct a rolling window of \(M\) consecutive macro-blocks with progressively increasing noise levels:
\begin{equation}
\mathcal{W}_k=\{B_k^{\sigma_1},B_{k+1}^{\sigma_2},\ldots,B_{k+M-1}^{\sigma_M}\},\,
\sigma_1<\cdots<\sigma_M .
\end{equation}
where \(B_i^{\sigma}\) denotes the \(i\)-th audio-video macro-block at noise level \(\sigma\). In each forward pass, all blocks inside \(\mathcal{W}_k\) are jointly denoised with bidirectional intra-modal and cross-modal attention within the window. The cleanest block, \( B_{k}^{\sigma_1}\), then exits the window, while the remaining blocks advance to the next denoising step with a new Gaussian-noise-corrupted block being appended. We train this stage with the same joint audio-video DMD objective as in Stage I, applied over all macro-blocks in \(\mathcal{W}_k\).

Because the active rolling window enables bidirectional interaction, earlier blocks are no longer frozen as potentially erroneous context for later blocks. Instead, they can leverage information from future blocks for refinement, thereby suppressing unidirectional error propagation. Moreover, audio latents form a one-dimensional temporal sequence and are easier to model than high-dimensional video latents, thus granting the video branch access to future audio cues therefore improves both the visual quality and cross-modal alignment.

\subsection{Synchronized Audio-Video Long-Short-Term Memory}

Existing streaming joint audio-video avatar generators~\citep{hallo-live} primarily rely on KV caches~\citep{zhao2025spatiavideogenerationupdatable} to preserve recent context during causal inference. Although a rolling KV cache can propagate visual and voice information across adjacent blocks, these attributes are inherited recursively from imperfect model predictions, so small deviations may accumulate over long horizons and eventually cause drift.

To counteract this drift, we introduce a synchronized audio-video long-short-term memory. We retain the KV states of the first audio-video macro-block \(B_0=(V_0,A_0)\) as long-term memory \(\mathcal{M}_{\mathrm{long}}\), which remains permanently visible to all subsequent blocks and provides a stable reference for identity, scene, voice timbre, and audio-video correspondence. In parallel, we maintain a rolling KV cache over the most recent \(L\) macro-blocks as short-term memory \(\mathcal{M}_{\mathrm{short}}\), which preserves motion and speech continuity. Given access to temporally aligned audio and video states, both memories achieve modality stability through self-attention and cross-modal fidelity through cross-attention.

As generation proceeds, the relative positional offset between the long-term memory and current queries increases, weakening the influence of the long-term memory embedded in the Rotary Position Embedding (RoPE). To address this problem, we periodically re-anchor the long-term memory. Specifically, We store its pre-RoPE keys \(\mathbf{K}^{\mathrm{orig}}\) and, after each five-second chunk has been generated, reapply RoPE to \(\mathbf{K}^{\mathrm{orig}}\) using positions immediately preceding the current rolling KV cache while keeping the corresponding values unchanged. This preserves the memory content while reducing the RoPE position gap between the long-term memory and current queries, maintaining the influence of the long-term memory during minute-level generation.

\subsection{Hierarchical Rolling Prompt Planning}

\begin{figure*}[!t]
  \centering
  \includegraphics[width=\textwidth]{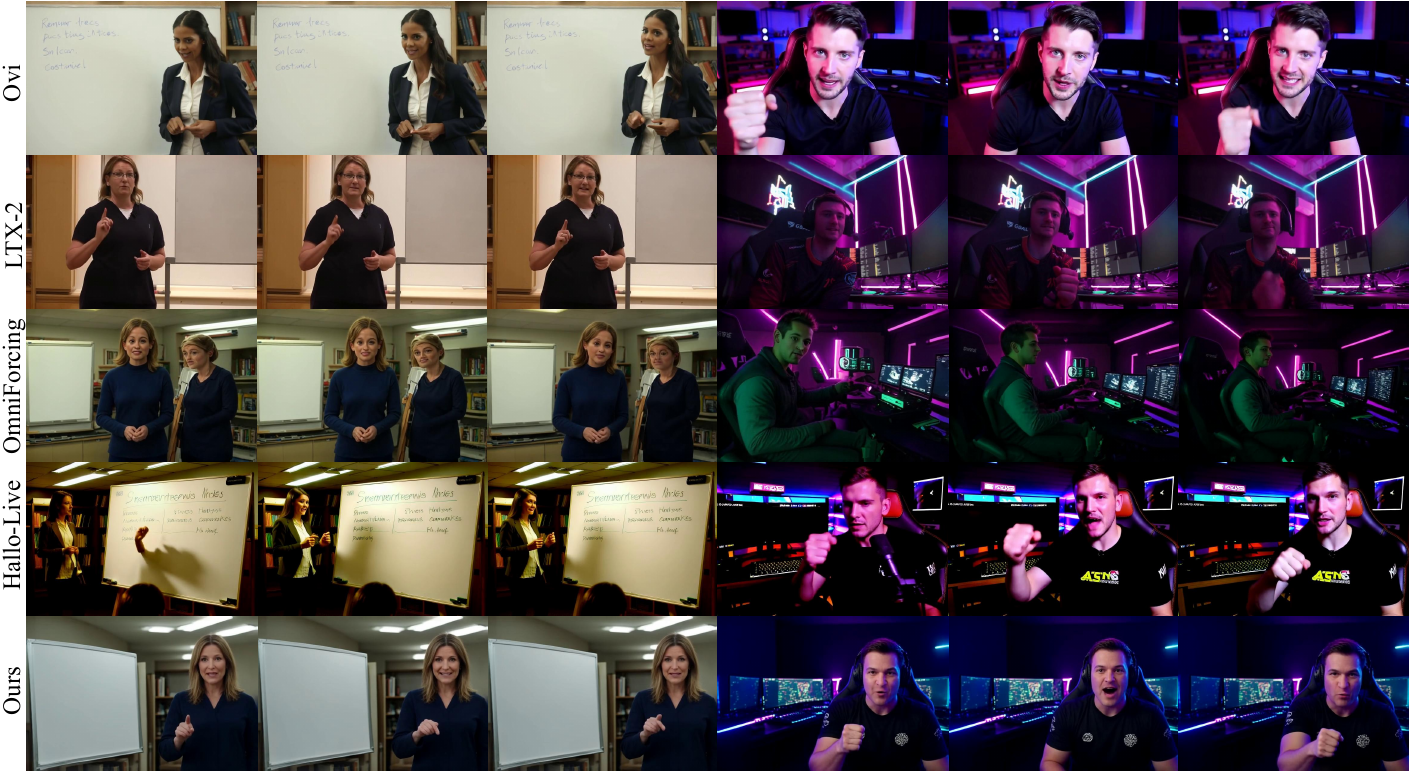}
  \caption{Qualitative comparison of 5-second avatar generation.
Omni-LiveAvatar generates realistic and temporally consistent avatars with visual quality comparable to Ovi and LTX-2, whereas OmniForcing exhibits noticeable realism degradation and Hallo-Live suffers from facial and hand artifacts as well as color drift.}
  \label{fig:qualitative_5s}
\end{figure*}



With progressive autoregressive distillation and synchronized audio-video long-short-term memory, Omni-LiveAvatar achieves stable real-time inference for long-horizon joint generation. However, the joint rolling-forcing paradigm introduces two new challenges for prompt planning. First, fixed-interval prompt switching, as adopted in LongLive~\citep{longlive}, is incompatible with rolling forcing: when a rolling window crosses a prompt boundary, blocks conditioned on different prompts are jointly denoised and attend to each other, causing semantic conflicts and leading to overlapping or blurred speech near transitions. Second, audio and video require textual conditioning at different temporal granularities. Visual semantics such as identity, scene, and motion evolve slowly and are better controlled by coarse-grained prompts, whereas speech content changes continuously and requires fine-grained sentence-level updates.

To address these challenges, we propose a hierarchical rolling prompt planning strategy. Specifically, we decompose the full prompt into a global prompt for persistent sequence-level semantics and local prompts for block-level speech content. For the global prompt \(\mathcal{P}^{\mathrm{global}}\), we keep its embedding fixed throughout generation, providing a stable global condition. For the local prompts, we partition the speech description into block-level segments, where \(\mathcal{P}^{\mathrm{local}}_{k}\) denotes the local prompt assigned to macro-block \(B_k\). For each rolling window \(\mathcal{W}_k\), the effective condition is formed by concatenating the fixed global prompt with the local prompts of the blocks within the rolling window:
\begin{equation}
\mathcal{P}(\mathcal{W}_k)
=
\left[
\mathcal{P}^{\mathrm{global}};\,
\mathcal{P}^{\mathrm{local}}_{k},
\ldots,
\mathcal{P}^{\mathrm{local}}_{k+M-1}
\right].
\end{equation}
When the rolling window advances from \(\mathcal{W}_k\) to \(\mathcal{W}_{k+1}\), the oldest local prompt \(\mathcal{P}^{\mathrm{local}}_{k}\) leaves the window and the local prompt \(\mathcal{P}^{\mathrm{local}}_{k+M}\) of the newly entering block is appended, so the effective condition shifts smoothly rather than switching abruptly at fixed boundaries. Together, the hierarchical rolling prompt planning enables smooth scheduling of textual conditions at different granularities under the joint rolling-forcing paradigm.

\section{Experiments}

\begin{table*}[t]
\centering
\small
\setlength{\tabcolsep}{3pt}
\renewcommand{\arraystretch}{1.15}
\resizebox{\textwidth}{!}{%
\begin{tabular}{@{}l|c|ccccc|c|ccc|ccc|c@{}}
\toprule
\multirow{2}{*}{Method}
& \multicolumn{1}{c|}{Speed}
& \multicolumn{5}{c|}{Video Quality}
& \multirow{2}{*}{VA$\uparrow$}
& \multicolumn{3}{c|}{Human Fidelity}
& \multicolumn{3}{c|}{Audio Quality}
& \multirow{2}{*}{Sync-C$\uparrow$} \\
\cmidrule(lr){2-2}
\cmidrule(lr){3-7}
\cmidrule(lr){9-11}
\cmidrule(lr){12-14}
& FPS$\uparrow$
& SC$\uparrow$
& BC$\uparrow$
& AQ$\uparrow$
& IQ$\uparrow$
& QS$\uparrow$
&
& HA$\uparrow$
& HI$\uparrow$
& HC$\uparrow$
& UT$\uparrow$
& DN$\uparrow$
& NI$\uparrow$
& \\
\midrule
\multicolumn{15}{l}{\emph{Bidirectional Models}} \\
LTX-2~\citep{LTX-2}
& 0.60
& 92.54 & 93.42 & 56.37 & 71.53 & 79.39
& 7.77
& 96.23 & 92.93 & 99.57
& 3.54 & 3.75 & 3.47
& 6.89 \\
Ovi~\citep{ovi}
& 1.41
& 97.10 & 95.29 & 33.13 & 65.83 & 76.14
& 6.51
& 98.55 & 97.13 & 99.19
& 3.10 & 3.72 & 2.63
& 6.88 \\
\midrule
\multicolumn{15}{l}{\emph{Real-time Autoregressive Models}} \\
OmniForcing~\citep{OmniForcing}
& 16.11
& 98.05 & 94.47 & 60.14 & 70.75 & 80.05
& 8.11
& 88.76 & 98.51 & \textbf{100.00}
& 2.46 & 3.84 & 3.03
& 1.60 \\
Hallo-Live~\citep{hallo-live}
& 16.50
& 98.20 & 93.64 & 27.97 & 66.04 & 74.21
& 8.04
& 98.39 & 97.42 & \textbf{100.00}
& 2.96 & 3.81 & 2.95
& 4.50 \\
\textbf{Omni-LiveAvatar}
& \textbf{19.57}
& \textbf{99.28} & \textbf{97.52} & \textbf{63.61}
& \textbf{72.11} & \textbf{81.72}
& \textbf{9.08}
& \textbf{98.76} & \textbf{100.00} & \textbf{100.00}
& \textbf{3.19} & \textbf{3.95} & \textbf{3.06}
& \textbf{6.16} \\
\bottomrule
\end{tabular}%
}
\caption{Quantitative comparison on 5-second avatar generation. The best results among the real-time autoregressive models are highlighted in bold, and \(\uparrow\) indicates that higher values are better.}
\label{tab:main_5s}
\end{table*}

\begin{table*}[t]
\centering
\small
\setlength{\tabcolsep}{3pt}
\renewcommand{\arraystretch}{1.15}
\resizebox{\textwidth}{!}{%
\begin{tabular}{@{}l|c|cccc|c|ccc|ccc|c@{}}
\toprule
\multirow{2}{*}{Method}
& \multicolumn{1}{c|}{Speed}
& \multicolumn{4}{c|}{Video Quality}
& \multirow{2}{*}{VA$\uparrow$}
& \multicolumn{3}{c|}{Human Fidelity}
& \multicolumn{3}{c|}{Audio Quality}
& \multirow{2}{*}{Sync-C$\uparrow$} \\
\cmidrule(lr){2-2}
\cmidrule(lr){3-6}
\cmidrule(lr){8-10}
\cmidrule(lr){11-13}
& FPS$\uparrow$
& SC$\uparrow$
& MS$\uparrow$
& AQ$\uparrow$
& IQ$\uparrow$
&
& HA$\uparrow$
& HI$\uparrow$
& HC$\uparrow$
& UT$\uparrow$
& DN$\uparrow$
& NI$\uparrow$
& \\
\midrule
\multicolumn{14}{l}{\emph{Real-time Autoregressive Models}} \\
OmniForcing~\citep{OmniForcing}
& 16.18
& 98.71 & 99.38 & 54.90 & 71.70
& 6.68
& 92.14 & 50.19 & \textbf{100.00}
& 1.60 & 3.51 & 2.90
& 0.28 \\
Hallo-Live~\citep{hallo-live}
& 13.80
& 99.50 & 99.19 & 47.16 & 61.18
& 5.46
& 97.81 & 67.60 & 77.20
& 2.02 & 3.97 & 2.99
& 0.72 \\
\textbf{Omni-LiveAvatar}
& \textbf{21.99}
& \textbf{99.55} & \textbf{99.68}
& \textbf{61.98} & \textbf{71.71}
& \textbf{9.82}
& \textbf{99.85} & \textbf{98.61} & \textbf{100.00}
& \textbf{2.80} & \textbf{4.04} & \textbf{3.24}
& \textbf{6.76} \\
\bottomrule
\end{tabular}%
}
\caption{Quantitative comparison on minute-level avatar generation. Omni-LiveAvatar achieves the best performance across all evaluated metrics, with particularly strong speed, aesthetic quality (AQ), human identity (HI) fidelity, and audio-video alignment.}
\label{tab:main_60s}
\end{table*}

In this section, we first describe the experimental setup, including the baselines, implementation details, and evaluation protocols in Section~\ref{sec:setup}. We then present the main results of the generation quality and speed for both 5-second and minute-level streaming generation in Section~\ref{sec:main}. Finally, we conduct ablation studies to validate the design choices of key components in Section~\ref{sec:ablation}.

\subsection{Experimental Setup}
\label{sec:setup}
\noindent\textbf{Baselines.}
We compare Omni-LiveAvatar against two representative frameworks for real-time autoregressive joint audio-video generation:
\begin{itemize}
\item \textbf{OmniForcing}~\citep{OmniForcing}: the current state-of-the-art (SOTA) framework for general text-to-audio-video (T2AV) generation.
\item \textbf{Hallo-Live}~\citep{hallo-live}: the current SOTA framework for avatar-specific audio-video generation.
\end{itemize}



\noindent\textbf{Metrics.}
To comprehensively evaluate Omni-LiveAvatar, we compare model capabilities across six dimensions: video quality, text-video alignment, human fidelity, audio quality, audio-video alignment, and generation speed.

For video quality (VQ), we use VBench~\citep{vbench} to evaluate per-frame fidelity and temporal coherence, reporting the Quality Score (QS) together with Subject Consistency (SC), Background Consistency (BC), Aesthetic Quality (AQ), and Imaging Quality (IQ). For long-form video quality, we use VBench-Long~\citep{vbenchlong} and report Subject Consistency (SC), Motion Smoothness (MS), Aesthetic Quality (AQ), and Imaging Quality (IQ). For text-video alignment, we use VideoAlign (VA)~\citep{videoalign} to assess semantic consistency between the generated video and the text prompt. For text-video alignment, we use VideoAlign (VA)~\citep{videoalign}. For human fidelity, we adopt human-centric metrics from VBench 2.0~\citep{vbench20} to evaluate the quality and consistency of generated avatars, including Human Anatomy (HA), Human Identity (HI), and Human Clothing (HC). For audio quality, we report three non-intrusive mean opinion score (MOS) predictors commonly used in speech synthesis and speech quality assessment: UTMOS~\citep{UTMOS} (UT) for speech naturalness, DNSMOS~\citep{DNSMOS} (DN) for noise and distortion, and NISQA~\citep{NISQA} (NI) for temporal continuity. For audio-video alignment, we use SyncNet confidence~\citep{syncnet} (Sync-C) to measure lip-speech synchronization. For generation speed, we follow Rolling Forcing~\citep{rollingforcing} and report throughput in frames per second (FPS) on a single NVIDIA H200 GPU to quantify real-time streaming capability.

\paragraph{Implementation Details.}
We build Omni-LiveAvatar on top of the 19B LTX-2 model~\citep{LTX-2}. All training is conducted on 8 NVIDIA H200 GPUs in bf16 precision using the AdamW optimizer with a global batch size of 8. We train Omni-LiveAvatar on an internal dataset of 35K text prompts. Using these prompts, we generate the Stage~II ODE regression pairs from our Stage-I few-step model. We train Stage~I, Stage~II, and Stage~III for 4{,}000, 3{,}000, and 3{,}000 steps, with learning rates of \(2\times10^{-5}\), \(1\times10^{-4}\), and \(2\times10^{-5}\), respectively. For the training objective, the video and audio losses are equally weighted (\(\lambda_v=\lambda_a=1.0\)) across all stages, and the classifier-free guidance scales in the DMD stages (Stage~I and Stage~III) are set to \(3\) and \(5\) for the video and audio streams, respectively. For the memory configuration, both the rolling-forcing window and the rolling KV cache span \(4\) macro-blocks, while the long-term memory retains the first macro-block. During inference, our few-step generator produces each block in four denoising steps at a resolution of \(512\times768\), enabling streaming generation up to minute-level duration.






\subsection{Main Results}
\label{sec:main}
We compare Omni-LiveAvatar against two families of baselines: offline bidirectional
joint audio-video generation models, including LTX-2~\citep{LTX-2} and
Ovi~\citep{ovi}, and real-time autoregressive distilled models, including
OmniForcing~\cite{OmniForcing} and Hallo-Live~\citep{hallo-live}. We conduct
comprehensive experiments on both 5-second and 60-second sequence generation, demonstrating that
Omni-LiveAvatar outperforms the autoregressive distilled baselines in both short-clip
and long-sequence generation.

\paragraph{Generation Results on 5-Second Clips.}
We present the quantitative and qualitative results in Table~\ref{tab:main_5s} and
Figure~\ref{fig:qualitative_5s}, respectively. As shown in Figure~\ref{fig:qualitative_5s},
OmniForcing~\citep{OmniForcing} produces avatars with degraded realism and Hallo-Live~\citep{hallo-live} suffers from facial artifacts, spurious hands, and noticeable color bias and drift. In contrast,
Omni-LiveAvatar generates natural and consistent avatars, achieving visual
quality comparable to the two bidirectional models. Quantitatively, Omni-LiveAvatar achieves real-time generation at 19.57 FPS on a single NVIDIA H200 GPU, delivering a \(\sim\)33\(\times\) speedup over its bidirectional teacher LTX-2
and a speedup of about 1.2\(\times\) over the autoregressive distilled baselines.
In terms of video quality and human-centric fidelity, Omni-LiveAvatar outperforms all
competitors, including both bidirectional and autoregressive models. Its overall quality
score (QS) exceeds the second-best result by 1.7~(81.72 vs.\ 80.05), and its human
identity (HI) and clothing (HC) fidelity both reach a perfect score of 100. Omni-LiveAvatar also
achieves the best text-video alignment (VA), surpassing the others by at least 0.97~(9.08 vs.\ 8.11). For audio quality, Omni-LiveAvatar consistently outperforms the autoregressive
baselines across all metrics, with only marginal
degradation in naturalness (UT) and continuity (NI) compared with the bidirectional teacher. A
similar trend is observed for audio-video synchronization (Sync-C). Overall, these results demonstrate that Omni-LiveAvatar achieves real-time generation without sacrificing  quality, consistently outperforming real-time baselines across all evaluated metrics while remaining competitive with offline bidirectional models in quality.

\paragraph{Generation Results on 60-Second Sequences.}
We show the quantitative and qualitative results for minute-level generation in Table~\ref{tab:main_60s} and
Figure~\ref{fig:teaser}, respectively. As shown in Figure~\ref{fig:teaser}, both the avatars and the backgrounds generated by OmniForcing and Hallo-Live exhibit drastic color and structural distortions, whereas Omni-LiveAvatar maintains stable and consistent visual appearance throughout the one-minute sequence. To visualize audio-video synchronization at the syllable level, we transcribe the generated speech using Whisper~\citep{whisper}, obtain syllable timestamps through forced alignment~\citep{forcealign}, and inspect the mouth shapes in the corresponding video frames. Figure~\ref{fig:teaser} shows that the lip motion generated by Omni-LiveAvatar remains closely aligned with the speech throughout the minute-level generation. Quantitatively, Omni-LiveAvatar outperforms the autoregressive-distilled baselines across all evaluated metrics. Specifically, it achieves real-time streaming generation at 21.99 FPS, running \(1.36\times\) faster than the second-fastest baseline (16.18 FPS). The quality improvements are particularly notable in human identity (HI, 98.61 vs.\ 67.60 and 50.19), audio-video synchronization (Sync-C, 6.76 vs.\ 0.72 and 0.28), and text-video alignment (VA, 9.82 vs.\ 6.68 and 5.46). These results demonstrate that Omni-LiveAvatar not only excels on short
clips but also pioneers real-time, high-fidelity, and temporally coherent
streaming generation at the minute scale.

\subsection{Ablation Study}
\label{sec:ablation}
\paragraph{Ablation on Training Methods.}
To validate our proposed progressive autoregressive distillation strategy, we perform ablations on joint rolling forcing (RF) and trajectory-aligned causal initialization (TA-ODE). Specifically, we compare our method against three settings: (1) using joint self forcing instead of
joint rolling forcing and generating the ODE pairs with the multi-step teacher
(w/o RF \& w/o TA-ODE); (2) using joint self forcing instead of joint rolling
forcing (w/o RF); and (3) generating the ODE pairs with the multi-step
teacher (w/o TA-ODE). As shown in
Table~\ref{tab:ablation_training}, the w/o RF \& w/o TA-ODE setting
degrades drastically across all evaluated quality metrics. Adding either rolling forcing or TA-ODE alone already yields
consistent improvements across all metrics over the w/o RF \& w/o TA-ODE variant. However, compared with ours, w/o RF mainly degrades audio quality and audio-video
synchronization, reducing UTMOS from $3.19$ to $1.93$ and Sync-C from
$6.16$ to $4.58$, whereas w/o TA-ODE primarily degrades
video quality and text-video alignment, with Qulity Score~(QS) dropping from $81.72$ to
$81.25$ and VideoAlign~(VA) from $9.08$ to $8.63$. Only when trajectory-aligned causal initialization and joint rolling forcing are
combined does the model achieve the best performance,
confirming that the two are jointly necessary.


\paragraph{Ablation on Memory Mechanism.}
To validate our synchronized audio-video long-short-term memory, we compare
it against three settings: (1) without the long-term memory
(w/o LTM); (2) without periodic RoPE
re-anchoring (w/o re-anchor); and (3) with temporally misaligned audio and
video long-term memory (async LTM).
As shown in Table~\ref{tab:ablation_memory}, the video branch is more dependent on the memory mechanism than the audio branch during long-horizon
inference, which indicates that video generation is more vulnerable to temporal drift due to its  modality complexity.
Specifically, w/o LTM causes the largest degradation in both video
quality and audio-video synchronization, with Aesthetic Quality dropping from
$61.98$ to $58.61$ and SyncNet confidence from $6.76$ to $5.79$. Removing re-anchoring mainly weakens the video quality metrics while async LTM primarily harms cross-modal alignment qualified by SyncNet confidence (Sync-C). These results
demonstrate that our synchronized audio-video long-short-term memory guarantees
the stability and audio-video alignment of long-horizon generation with sophisticated long-term memory and re-anchoring designs.


\paragraph{Ablation on Prompt Planning Strategy.}
To validate our hierarchical rolling prompt planning, we compare it against two
settings: (1) providing the full prompt to the model at once
(w/o rolling prompt); and (2) updating the prompt every 5\,s
(5\,s-interval switching).
As shown in Table~\ref{tab:ablation_prompt}, the audio-related metrics are more
sensitive to the prompt planning strategy than the video-related metrics,
indicating that audio requires a finer-grained and smoother prompt schedule.
Specifically, exerting the full prompt at once yields the largest drops across nearly all metrics, notably including the audio naturalness (UT) from $2.80$ to $1.60$ and SyncNet confidence (Sync-C) from $6.76$ to $4.47$. Updating the prompt every 5\,s mainly degrades the audio-quality metrics as well as audio-video alignment, with SyncNet
confidence (Sync-C) dropping from $6.76$ to $5.11$. These results show that, under joint rolling-forcing inference paradigm, our
hierarchical rolling prompt planning enables smooth, continuous, and aligned audio-video generation

\begin{table}[t]
\centering
\footnotesize
\setlength{\tabcolsep}{2pt}
\renewcommand{\arraystretch}{1.25}
\begin{tabular}{@{}l@{\hspace{5pt}}c@{\hspace{3pt}}ccc@{\hspace{3pt}}c@{\hspace{3pt}}c@{}}
\toprule
\multirow{2}{*}{Variant}
& \multicolumn{1}{c}{VQ}
& \multicolumn{3}{c}{Audio Quality}
& \multirow{2}{*}{Sync-C$\uparrow$}
& \multirow{2}{*}{VA$\uparrow$} \\
\cmidrule(lr){2-2}
\cmidrule(lr){3-5}
& QS$\uparrow$
& UT$\uparrow$
& DN$\uparrow$
& NI$\uparrow$
& & \\
\midrule
w/o RF \& w/o TA-ODE
& 72.05 & 1.51 & 3.21 & 2.70 & 1.23 & 1.24 \\
w/o RF
& 80.67 & 1.93 & 3.38 & 2.44 & 4.58 & 8.09 \\
w/o TA-ODE
& 81.25 & 3.14 & 3.81 & 2.96 & 5.98 & 8.63 \\
\textbf{Ours}
& \textbf{81.72} & \textbf{3.19} & \textbf{3.95}
& \textbf{3.06} & \textbf{6.16} & \textbf{9.08} \\
\bottomrule
\end{tabular}
\caption{Ablation of training strategies.}
\label{tab:ablation_training}
\end{table}


\begin{table}[t]
\centering
\footnotesize
\setlength{\tabcolsep}{2pt}
\renewcommand{\arraystretch}{1.25}
\begin{tabular}{@{}l@{\hspace{5pt}}cc@{\hspace{3pt}}ccc@{\hspace{3pt}}c@{\hspace{3pt}}c@{}}
\toprule
\multirow{2}{*}{Variant}
& \multicolumn{2}{c}{Video Quality}
& \multicolumn{3}{c}{Audio Quality}
& \multirow{2}{*}{Sync-C$\uparrow$}
& \multirow{2}{*}{VA$\uparrow$} \\
\cmidrule(lr){2-3}
\cmidrule(lr){4-6}
& AQ$\uparrow$
& IQ$\uparrow$
& UT$\uparrow$
& DN$\uparrow$
& NI$\uparrow$
& & \\
\midrule
w/o LTM
& 58.61 & 71.16 & 2.79 & 4.03 & 3.22 & 5.79 & 9.70 \\
w/o re-anchor
& 61.16 & 71.43 & 2.78 & 4.02 & 3.23 & 6.70 & 9.82 \\
async LTM
& 61.93 & 71.69 & 2.76 & 4.03 & 3.21 & 6.61 & 9.80 \\
\textbf{Ours}
& \textbf{61.98} & \textbf{71.71}
& \textbf{2.80} & \textbf{4.04} & \textbf{3.24}
& \textbf{6.76} & \textbf{9.82} \\
\bottomrule
\end{tabular}
\caption{Ablation of memory mechanisms.}
\label{tab:ablation_memory}
\end{table}

\begin{table}[!t]
\centering
\footnotesize
\setlength{\tabcolsep}{1.5pt}
\renewcommand{\arraystretch}{1.25}
\begin{tabular}{@{}l@{\hspace{3pt}}cc@{\hspace{2pt}}ccc@{\hspace{2pt}}c@{\hspace{2pt}}c@{}}
\toprule
\multirow{2}{*}{Variant}
& \multicolumn{2}{c}{VQ}
& \multicolumn{3}{c}{Audio Quality}
& \multirow{2}{*}{Sync-C$\uparrow$}
& \multirow{2}{*}{VA$\uparrow$} \\
\cmidrule(lr){2-3}
\cmidrule(lr){4-6}
& AQ$\uparrow$
& IQ$\uparrow$
& UT$\uparrow$
& DN$\uparrow$
& NI$\uparrow$
& & \\
\midrule
w/o rolling prompt
& 55.74 & 68.45 & 1.60 & 3.28 & 2.38 & 4.47 & 9.72 \\
5\,s-interval switching
& 61.23 & 71.33 & 2.65 & 3.94 & 3.19 & 5.11 & 9.71 \\
\textbf{Ours}
& \textbf{61.98} & \textbf{71.71}
& \textbf{2.80} & \textbf{4.04} & \textbf{3.24}
& \textbf{6.76} & \textbf{9.82} \\
\bottomrule
\end{tabular}
\caption{Ablation of prompt planning strategies.}
\label{tab:ablation_prompt}
\end{table}

\section*{Conclusions}

In this paper, we present Omni-LiveAvatar, the first framework for minute-level, real-time streaming joint audio-video avatar generation. We introduce a progressive autoregressive distillation pipeline to transform a large bidirectional audio-video diffusion model into a few-step causal generator; a synchronized audio-video long-short-term memory to preserve long-duration consistency and cross-modal alignment under a bounded memory budget; and a hierarchical rolling prompt planning strategy to enable coherent semantic evolution and seamless prompt transitions under the rolling-forcing setting. Extensive experiments demonstrate that Omni-LiveAvatar achieves a \(33\times\) speedup over its LTX-2 teacher on a single NVIDIA H200 GPU while outperforming real-time baselines across all evaluated metrics.
\bigskip

\bibliography{AnonymousSubmission2027}

\clearpage
\newpage
\appendix

\onecolumn

\begin{center}
    {\Large\bfseries
    Additional Qualitative Comparisons on 60-Second Avatar Generation}
\end{center}
\vspace{0.6em}

\begin{center}
    \includegraphics[width=\textwidth]{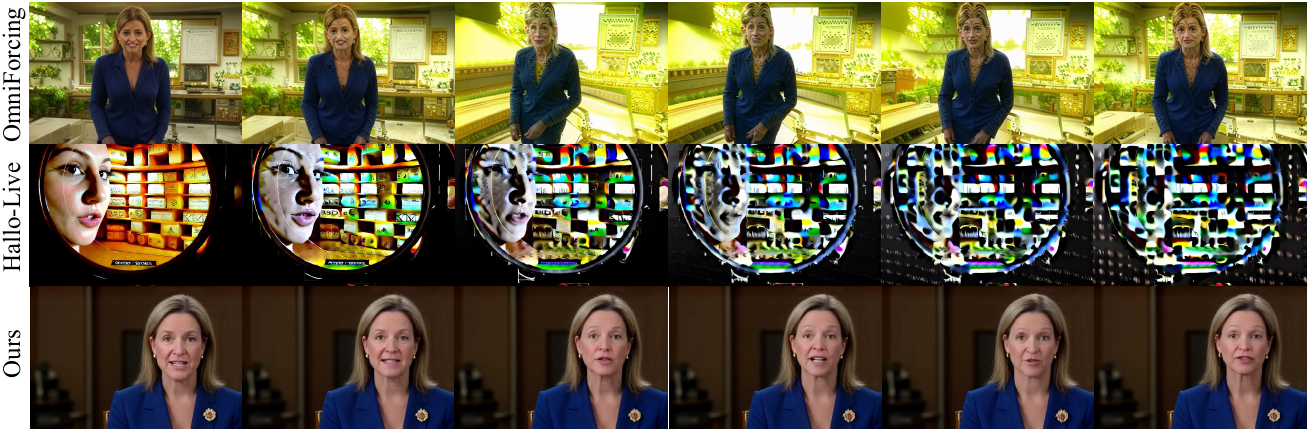}\par
    \smallskip

    \includegraphics[width=\textwidth]{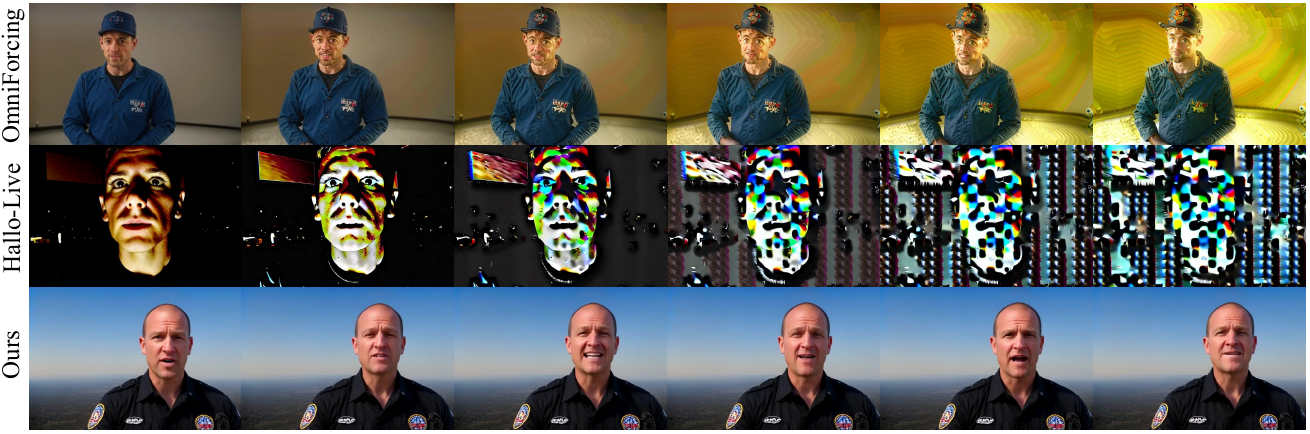}\par
    \smallskip

    \includegraphics[width=\textwidth]{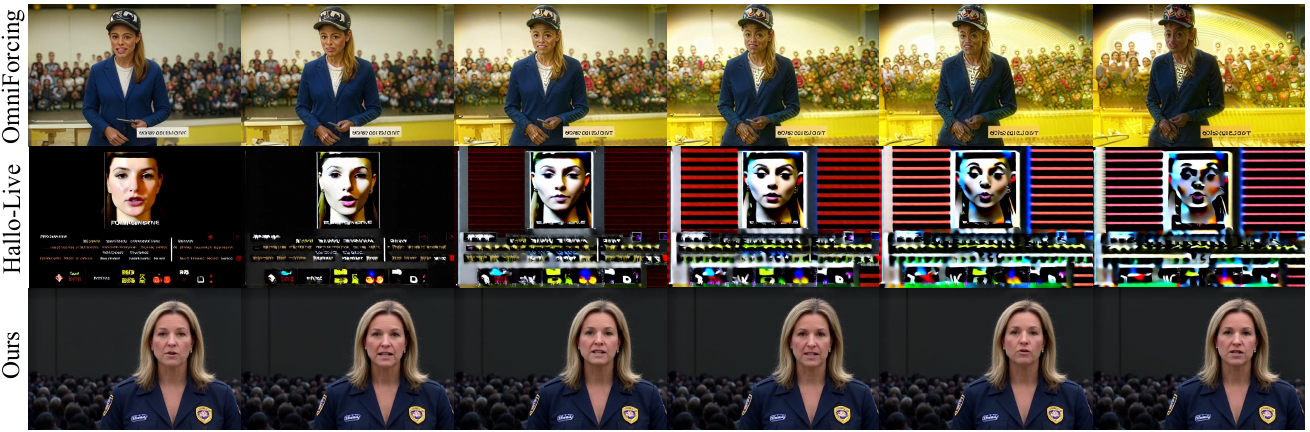}\par

    \captionof{figure}{Additional qualitative comparisons on 60-second avatar
    generation. Omni-LiveAvatar maintains consistent avatar and background
    appearance and accurate audio-video synchronization throughout minute-long
    generation.}
    \label{fig:media_supp}
\end{center}

\clearpage

\begin{center}
    \includegraphics[width=\textwidth]{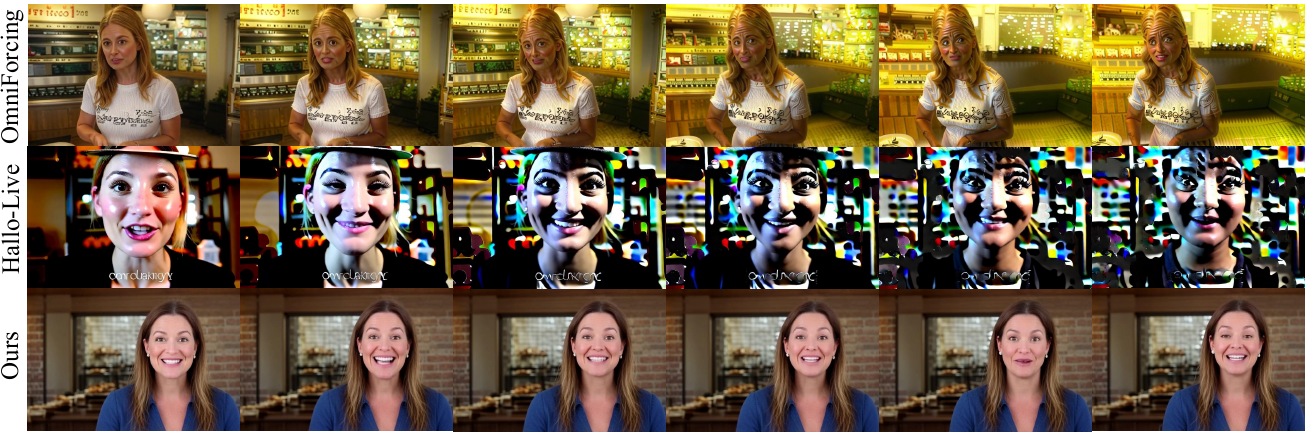}\par
    \smallskip

    \includegraphics[width=\textwidth]{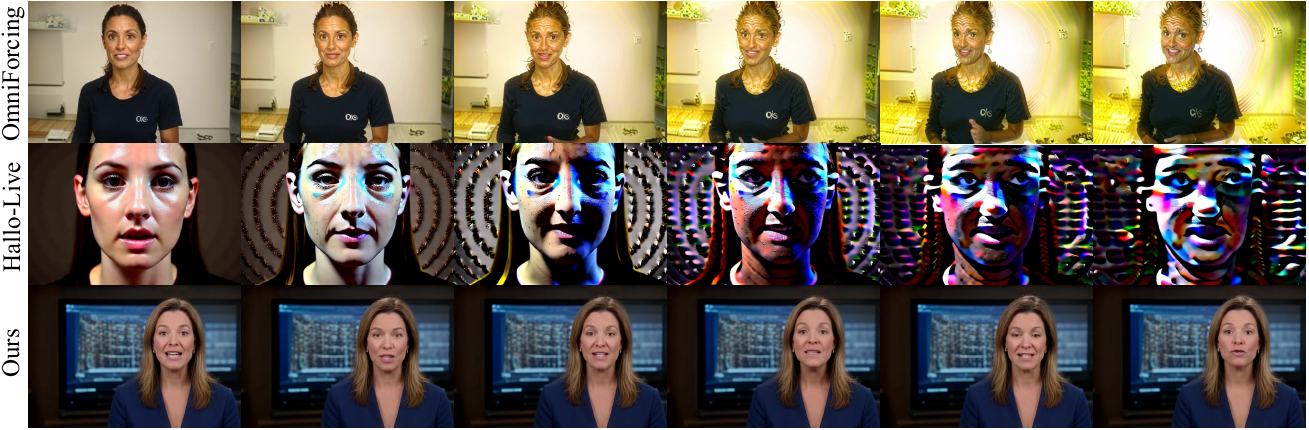}\par
    \smallskip

    \includegraphics[width=\textwidth]{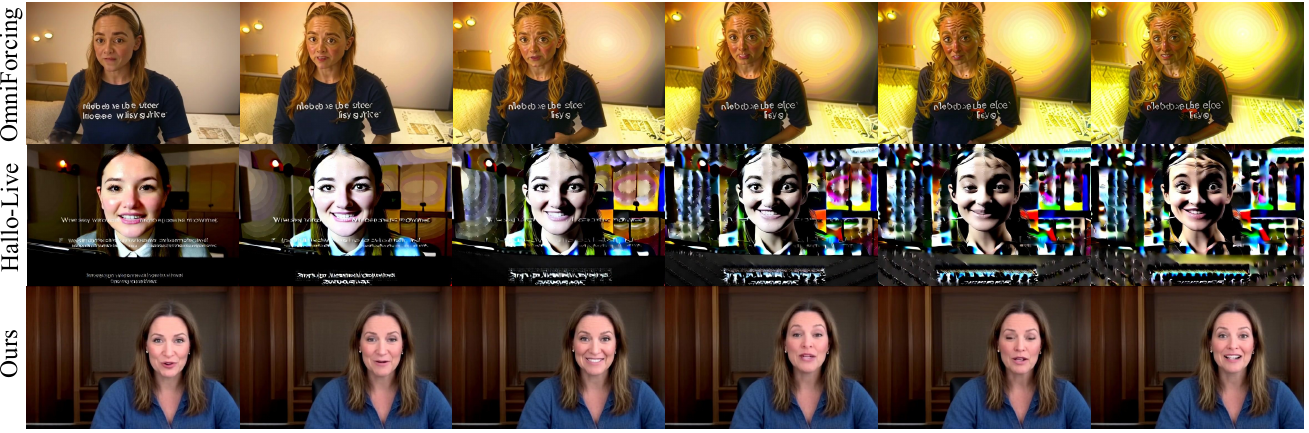}\par

    \captionof*{figure}{Figure~\ref{fig:media_supp} (continued).}
\end{center}

\end{document}